\documentclass[pdflatex,sn-mathphys-num]{sn-jnl}

\usepackage{graphicx}%
\usepackage{multirow}%
\usepackage{amsmath,amssymb,amsfonts}%
\usepackage{amsthm}%
\usepackage{mathrsfs}%
\usepackage[title]{appendix}%
\usepackage{xcolor}%
\usepackage{textcomp}%
\usepackage{manyfoot}%
\usepackage{booktabs}%
\usepackage{algorithm}%
\usepackage{algorithmicx}%
\usepackage{algpseudocode}%
\usepackage{listings}%

\theoremstyle{thmstyleone}%
\theoremstyle{thmstyletwo}%

\theoremstyle{thmstylethree}%

\begin{document}

\title[Towards High-Fidelity and Controllable Bioacoustic Generation via Enhanced Diffusion Learning]{Towards High-Fidelity and Controllable Bioacoustic Generation via Enhanced Diffusion Learning}


\author[1]{\fnm{Tianyu} \sur{Song}}

\author*[1]{\fnm{Ton Viet} \sur{Ta}}

\affil*[1]{\orgdiv{Graduate School of Bioresource and Bioenvironmental Science}, \orgname{Kyushu University}, \orgaddress{\street{744 Motooka, Nishi Ward}, \city{Fukuoka }, \postcode{819-0395}, \country{Japan}}}



\abstract{
Generative modeling offers new opportunities for bioacoustics, enabling the synthesis of realistic animal vocalizations that could support biomonitoring efforts and supplement scarce data for endangered species. However, directly generating bird call waveforms from noisy field recordings remains a major challenge.

We propose \textit{BirdDiff}, a generative framework designed to synthesize bird calls from a noisy dataset of 12 wild bird species. The model incorporates a ``zeroth layer'' stage for multi-scale adaptive bird-call enhancement, followed by a diffusion-based generator conditioned on three modalities: Mel-frequency cepstral coefficients, species labels, and textual descriptions. The enhancement stage improves signal-to-noise ratio (SNR) while minimizing spectral distortion, achieving the highest SNR gain (+10.45\,dB) and lowest Itakura–Saito Distance (0.54) compared to three widely used non-training enhancement methods.

We evaluate BirdDiff against a baseline generative model, DiffWave. Our method yields substantial improvements in generative quality metrics: Fréchet Audio Distance (0.590~$\rightarrow$~0.213), Jensen–Shannon Divergence (0.259~$\rightarrow$~0.226), and Number of Statistically-Different Bins (7.33~$\rightarrow$~5.58). To assess species-specific detail preservation, we use a ResNet50 classifier trained on the original dataset to identify generated samples. Classification accuracy improves from 35.9\% (DiffWave) to 70.1\% (BirdDiff), with 8 of 12 species exceeding 70\% accuracy.

These results demonstrate that BirdDiff enables high-fidelity, controllable bird call generation directly from noisy field recordings.}





\maketitle

\section{Introduction}
Acoustic sensing technologies have become essential tools for ecological monitoring~\cite{van2023automated}, enabling the identification and tracking of categories through their vocalizations. Among these, bird calls are particularly informative: they carry detailed cues about categories identity, behavior, and environmental context. However, in practice, bird call datasets are often limited in scope and quality. Field recordings are frequently corrupted by background noise, and the scarcity of labeled data—especially for rare or elusive categories—hampers both ecological research and the development of automated categories recognition systems~\cite{song2024,song2025}.

In parallel, generative models have rapidly advanced across multiple modalities, including vision, text, and audio~\cite{goodfellow2014generative,tian2024visual,hoogeboom2021autoregressive}. Diffusion models~\cite{ho2020denoising}, initially developed for high-fidelity image generation, have recently shown promising performance in audio synthesis tasks. For example, they have been applied to speech enhancement~\cite{lemercier2024diffusion}, semantic correction using large language models~\cite{huang2025audio}, and spatial audio generation~\cite{kushwaha2025diff}. Diff-SAGe~\cite{kushwaha2025diff} integrates diffusion processes with transformers to synthesize immersive soundscapes, while Grassucci et al.~\cite{grassucci2024diff} demonstrated how conditional diffusion can enable semantic audio generation.

Despite these advances, the application of generative models in bioacoustics remains limited. Most prior work focuses on denoising or data augmentation rather than direct waveform synthesis. For example, Herbst et al.~\cite{herbst2024empirical} compared denoising diffusion models and variational autoencoders (VAEs) for augmenting primate vocalizations, finding that traditional enhancement methods can sometimes match the performance of generative approaches. In the context of birds, Kumar et al.~\cite{kumar2024vision} and Zhang et al.~\cite{zhang2023birdsounds} used visual-audio techniques to isolate calls from noisy recordings. Meanwhile, other studies have explored cross-modal synthesis such as generating bird images from sound~\cite{shim2021s2i} or generating bird calls from visual inputs~\cite{guei2024ecogen}, often by producing spectrograms that are then converted to waveforms.

While spectrogram-based generation is effective, converting between waveforms and spectrograms—particularly via the short-time Fourier transform (STFT)—can lead to information loss, especially in phase reconstruction. Direct waveform generation avoids these limitations and preserves more of the original acoustic characteristics. However, high background noise in bird call recordings poses serious challenges for waveform-based models such as DiffWave~\cite{kong2020diffwave} and WaveNet~\cite{oord2016wavenet}, which often fail under such conditions and produce audio lacking realism or intelligibility.

To address these limitations,  we propose \textit{BirdDiff}, a generative framework for synthesizing bird calls from a noisy dataset of 12 wild bird species. The model incorporates a ``zeroth layer'' stage for multi-scale adaptive bird-call enhancement, followed by a diffusion-based generator conditioned on three modalities: Mel-frequency cepstral coefficients (MFCC), species labels, and textual descriptions. 

The ``zeroth layer'' stage emphasizes amplifying category-specific acoustic cues to help the generator distinguish bird calls from background interference, rather than attempting to completely remove noise. It combines spectral subtraction~\cite{boll1979spectral} with multi-band fusion, an effective variant of spectral subtraction~\cite{upadhyay2015speech}. This approach has evolved with advances such as adaptive noise estimation for speech activity detection~\cite{upadhyay2013improved}, hybrid architectures with discrete wavelet transforms~\cite{iqbal2024dwt}, and neural network–based adaptive noise estimation~\cite{liu2025acoustic}. Our contribution introduces frequency-band weighting to avoid unnecessary spectral subtraction in clean regions, along with intelligent noise selection to apply subtraction only where noise dominates. Applied to the noisy dataset of 12 bird species, this stage alone improves signal-to-noise ratio (SNR) by +10.45\,dB and achieves the lowest Itakura–Saito Distance (0.54) compared to three widely used non-training enhancement methods~\cite{gray1980distortion,basseville1989distance}.

The next stage of BirdDiff integrates spectral features, species labels, and textual descriptions to guide the diffusion-based generator in producing category-specific bird calls. Compared with the baseline DiffWave model, BirdDiff demonstrates substantial improvements in generative quality: Fréchet Audio Distance (FAD)~\cite{kilgour2018fr} decreases from 0.590 to 0.213, Jensen–Shannon Divergence (JSD)~\cite{menendez1997jensen} from 0.259 to 0.226, and the Number of Statistically-Different Bins (NDB)~\cite{richardson2018gans} from 7.33 to 5.58 (these metrics are defined in the next section). To evaluate species-specific detail preservation, we use a ResNet50 classifier trained on the original dataset to identify generated samples. Classification accuracy rises from 35.9\% for DiffWave to 70.1\% for BirdDiff, with 8 of the 12 species achieving over 70\% accuracy.

Finally, while previous studies such as NatureLM-Audio~\cite{robinson2024naturelm} have adapted foundation models to bioacoustics primarily for classification, our work extends this line of research by directly generating bird call waveforms with category and semantic control, thereby enabling waveform-level synthesis tailored to ecological applications. We believe this contribution offers a practical and scalable solution for biodiversity assessment, categories simulation, and future ecological research.

The remainder of this paper is organized as follows. Section~\ref{Materials} describes the original dataset and presents an overview of the proposed BirdDiff framework. Section~\ref{results} reports the experimental results and evaluations. Finally, Section~\ref{Conclusion} concludes the paper and discusses potential directions for future research.

\section{Materials and Experimental Design} \label{Materials}
In this section, we first describe the original dataset in Subsection~\ref{dataset}. We then introduce the zeroth layer of our BirdDiff model in Subsection~\ref{zerothlayer}. Subsection~\ref{MainGenerator} presents the architecture of the main generator of BirdDiff. Finally, Subsection~\ref{EvaluationMethodology} outlines the metrics used to evaluate our model.

\subsection{Dataset} \label{dataset}
The dataset used in this study is provided by Bird Data Technology (Beijing) Co., Ltd.~\cite{BD_2023}. It consists of standardized natural sound recordings that have been manually annotated and validated by domain experts. We utilize   a subset containing 12 bird categories, including both single-species and multi-species groups:
\begin{itemize}
    \item Single-species categories: Mallard, Red-throated Diver, Grey Heron, Common Buzzard, Western Water Rail, Woodcock, Bar-tailed Godwit
    \item Multi-species categories: Teal, Quail, Pheasant, Redshank, Sparrow
\end{itemize}
For simplicity, we refer to each of these 12 categories as a ``species'' throughout this paper.  
In total, the dataset includes 6,610 bird call audio clips, each 2 seconds in duration. To ensure consistency in subsequent processing, all recordings were converted to single-channel WAV format with a sampling rate of 22.05\,kHz and a bitrate of 705.9\,kbps.

\subsection{Zeroth Layer: Adaptive Enhancement Stage} \label{zerothlayer}
The zeroth layer of BirdDiff serves as an adaptive enhancement stage, designed to amplify category-specific acoustic cues and thereby help the generator distinguish bird calls from background interference. This stage combines spectral subtraction with multi-band fusion, a variant that allows selective noise reduction across frequency bands. In particular, we introduce frequency-band weighting to avoid unnecessary subtraction in relatively clean regions and employ intelligent noise selection that applies subtraction only in noise-dominated areas.

We begin by estimating the signal-to-noise ratio (SNR) of the audio recordings. Following the Segmental SNR (SegSNR) metric~\cite{plapous2006improved}, we compute the SNR in decibel scale as
\[
\text{SegSNR} = \frac{1}{M} \sum_{m=0}^{M-1} 10 \log_{10} \left( \frac{\sum_{n=0}^{L-1} s_{\mathrm{est}, m}^2(n)}{\sum_{n=0}^{L-1} n_{\mathrm{est}, m}^2(n) + \epsilon} \right),
\]
where $s_{\mathrm{est}}$ and $n_{\mathrm{est}}$ denote estimates of the signal and noise components, respectively. In practice, we approximate these components directly from the raw audio: $s_{\mathrm{est}}$ is obtained by applying a band-pass filter between 2--8~kHz to the waveform $x$, and the residual is taken as the noise estimate $n_{\mathrm{est}} = x - s_{\mathrm{est}}$.  

Many samples exhibit negative SNR values, indicating that bird vocalizations are heavily masked by background noise. For diffusion models, low-SNR inputs can cause degeneration, where the model inadvertently learns noise distributions rather than meaningful signal patterns. Consequently, preprocessing to improve SNR is essential.

Rather than developing a full denoiser, we propose a lightweight, adaptive multi-band enhancement tailored to improve the SNR while preserving the intrinsic frequency characteristics of bird calls. Our approach uses multi-scale frequency decomposition to isolate bird call components across different frequency bands. Each band is assigned an adaptive weight $w_i$ based on energy distribution and spectral relevance. These weights are used to reconstruct an enhanced signal $s_{\text{est}}$ and residual noise $r_{\text{rn}}$, while the noise reduction strength $\alpha'$ is dynamically adjusted according to the estimated SNR to ensure optimal enhancement under varying noise conditions.  

A key component of the method is residual noise selection: the most representative noise fragment $n_{\text{ref}}$ is automatically identified from the residual signal and then used in a spectral subtraction process. This strategy preserves critical bird call components while selectively reducing background noise. To the best of our knowledge, this is the first adaptive enhancement technique specifically tailored for bird call audio as a preprocessing step for diffusion models. In our implementation, the multi-band decomposition $\mathcal{B}$ consists of four overlapping frequency bands: (1500, 3000), (2500, 5000), (4000, 8000), and (7000, 11000) Hz. These ranges were selected based on extensive experiments with bird calls in our dataset, and are designed to capture both low–mid vocalizations and high-frequency components that are critical for species identification. The overall procedure is summarized in Algorithm~\ref{alg:msabe}.

\begin{algorithm}[htbp]
\caption{Multi-band Adaptive Bird-Call Enhancement}
\label{alg:msabe}
\begin{algorithmic}[1]

\Statex \textbf{Multi-Band Decomposition and Adaptive Weighting}
\For{each band $(f_{\text{low}}, f_{\text{high}}) \in \mathcal{B}$}
    \State $b_i \gets \text{BandpassFilter}(x, [f_{\text{low}}, f_{\text{high}}])$
    \State $w_i \gets \text{AdaptiveWeight}(b_i, x)$
\EndFor

\Statex \textbf{Weighted Signal Reconstruction}
\State $s_{\text{est}} \gets \sum_i \frac{w_i}{\sum_j w_j} \cdot b_i$
\State $r_{\text{rn}} \gets x - s_{\text{est}}$

\Statex \textbf{Adaptive Noise Reduction}
\State $\text{SNR}_{est} \gets \text{SegSNR}(s_{est},r_{rn})$
\State $\alpha' \gets \text{AdaptStrength}(\text{SNR}_{\text{est}})$
\State $n_{\text{ref}} \gets \text{SelectNoiseReference}(r_{\text{rn}})$
\State $r_{\text{clean}} \gets \text{SpectralSubtraction }(r_{\text{rn}}, n_{\text{ref}}, \alpha')$

\Statex \textbf{Final Reconstruction}
\State $x_{\text{enh}} \gets \text{Normalize}(s_{\text{est}} + r_{\text{clean}})$

\State \Return $(x_{\text{enh}}, \{w_i\})$
\end{algorithmic}
\end{algorithm}

\paragraph{Residual Denoising} 
Traditional full-spectrum denoising methods, such as spectral subtraction, operate directly on the entire frequency spectrum of the input signal. While these techniques can be effective for general noise reduction, they present notable limitations when applied to bird-call enhancement:
\begin{itemize}
    \item [(a)] When bird vocalizations and background noise occupy overlapping frequency ranges, traditional approaches struggle to separate them. This often leads to either incomplete noise removal or the inadvertent loss of critical bird call information.
    \item [(b)] Applying the same denoising process uniformly across all frequency bands can unintentionally suppress high-frequency components or other perceptually salient details in the bird calls.
\end{itemize}

To overcome these issues, we introduce a residual-domain processing paradigm that fundamentally addresses the limitations of full-spectrum approaches. As outlined in Step 6 of Algorithm~\ref{alg:msabe}, our method decomposes the input signal into multiple frequency bands, prioritizing the preservation of those that are most relevant to bird vocalizations.

The core idea is to isolate and retain perceptually important signal components via multi-scale bandpass filtering, while directing spectral subtraction exclusively to the residual components—those deemed less critical or dominated by noise. This selective strategy ensures that noise reduction is concentrated where it is most needed, without compromising the integrity of the original signal.

By relocating most of the noise energy to the residual domain and excluding key signal bands from aggressive processing, our approach significantly reduces typical spectral subtraction artifacts. As a result, bird calls retain their clarity and fidelity, even under challenging noise conditions.

\begin{figure*}[htbp]
\centerline{\includegraphics[width=\columnwidth]{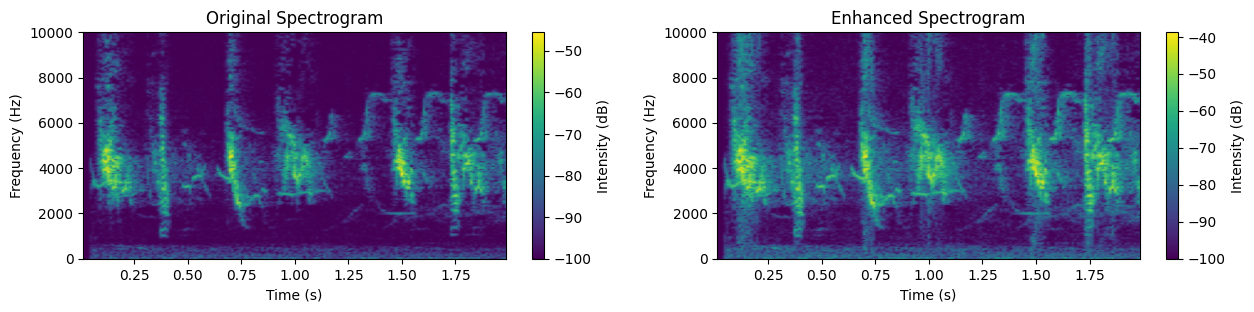}}
\caption{Illustration of enhancement result}
\label{fig:enhanced}
\end{figure*}

\paragraph{Effect verification}
Figure~\ref{fig:enhanced} illustrates the enhancement outcome. While SNR improvements (typically 3–15~dB) are measurable, the primary objective is to provide cleaner, structured inputs for the diffusion model, enabling it to learn accurate bird call distributions. Beyond numerical gains, the method preserves essential acoustic features—spectral shape, energy distribution, and phase continuity—often degraded by conventional denoising techniques. 

This targeted processing converts low-SNR recordings into clearer representations, supporting more effective model training and higher-quality bird call synthesis. Importantly, the method is task-specific, lightweight, and optimized for generative modeling, not general-purpose denoising.

In summary, our multi-scale adaptive enhancement framework combines bandpass filtering, adaptive weighting, and residual-domain spectral subtraction to provide a principled preprocessing strategy. This approach ensures effective noise reduction while preserving class-specific spectral cues, facilitating stable and accurate training of diffusion models for bird call generation.




\subsection{Main Generator: Diffusion-based Model with Multimodal Conditioning} \label{MainGenerator}   
Our model is built upon  DiffWave~\cite{kong2020diffwave}, a non-autoregressive diffusion model designed to generate raw audio waveforms. When trained on our adaptively enhanced dataset, this architecture successfully synthesizes bird vocalizations that  are clearly distinguishable across species.  In particular, the generated calls preserve the characteristic frequency patterns and acoustic signatures of each category, owing to the multimodal conditioning mechanism described below.  This mechanism integrates category labels, spectral features, and textual descriptions to steer the generation process toward the desired bird class.  In contrast, models trained on unprocessed data consistently fail to produce intelligible or ecologically valid calls, highlighting the critical role of data enhancement. 

\begin{figure}[htbp]
\centerline{\includegraphics[width=\columnwidth]{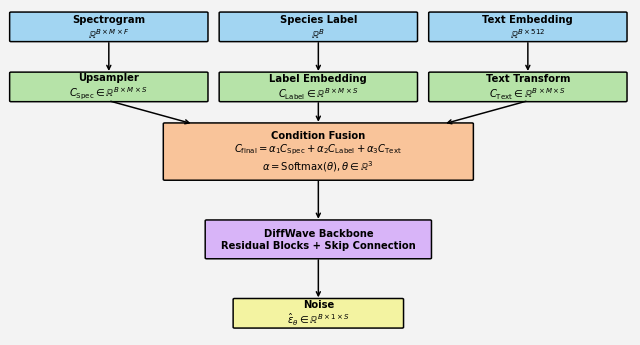}}
\caption{
Overview of the proposed architecture. Spectrogram, categories labels, and text embeddings are independently encoded and then fused via a learnable weighted sum to form the final condition vector. This fused representation conditions a DiffWave backbone to generate bird call waveforms with categories-level control and enhanced fidelity. The model learns to denoise a Gaussian noise input $\epsilon_0$ into clean audio signals.
}
\label{fig:str}
\end{figure}

To extend the baseline DiffWave for categories-controllable generation, we introduce multimodal conditioning mechanisms. As illustrated in Figure~\ref{fig:str}, the architecture accepts three types of conditioning information: Mel-frequency cepstral coefficients (MFCCs), categories labels, and textual descriptions. These inputs are encoded separately and then fused through a weighted attention mechanism to guide the generation process.

\paragraph{Spectrogram} 
Following the original DiffWave setup, we use MFCCs to represent spectral content, serving as the core acoustic conditioning.

\paragraph{Categories Labels} 
Each bird categories is assigned a learnable embedding vector. This enables class-level control and ensures the generated waveform matches the vocal characteristics of the target categories.

\paragraph{Textual Descriptions}
Free-form textual descriptions are encoded and projected into the same latent space as the MFCC features. This allows the model to incorporate semantic nuance and capture intra-categories variability.  
For each species, we use concise descriptive phrases emphasizing both acoustic and ecological characteristics. A consistent description was applied across all samples of the same species to provide stable semantic conditioning during generation. The descriptions for each species are as follows:

Mallard: Known for its lively equack, featuring bright, clear notes interspersed with moderate, slightly raspy undertones, conveying a brisk, familiar waterfowl sound.

Teal: A small, agile duck whose short, high-pitched whistles are rapid and delicate, reflecting a swift and spirited movement over marshes or ponds.

Quail: Characterized by a distinctive, concise ewet-my-lips or throbbing chirp, typically gentle yet sprightly, suggesting a lively presence in grasslands.

Pheasant: Issues a loud, abrupt crowing call that starts sharp, often followed by a drumming of wings, highlighting a bold, earthy resonance.

Red-throated Diver: Exhibits an eerie, wailing call with a blend of low, rumbling tones and high, drawn-out notes, suggesting a wild, misty lakeside atmosphere.

Grey Heron: Emits a low, harsh croak or guttural squawk, occasionally rattling, reflecting a deliberate and solitary presence along shorelines.

Common Buzzard: Utters a distinctive mewing epee-yow, moderate in pitch and plaintive, symbolizing open skies and rolling countryside.

Western Water Rail: Soft, squealing or grunting sounds with a faint pig-like timbre, discreet yet rhythmic, befitting a secretive marsh inhabitant.

Woodcock: Characterized by a low, froglike croak interspersed with a sharp, nasal esqueak, especially during dusk display flights, adding mystique to forest clearings.

Bar-tailed Godwit: Characterized by gentle, trilling ekoo-lik or ewit-wit calls, light and flowing, mirroring its long migratory flights and coastal stopovers.

Redshank: Noted for brisk, fluty eteu-teu notes, moderately pitched with a lively cadence, echoing across tidal flats or shallow, muddy waters.

Sparrow: A bright, chirpy series of quick notes, lively and conversational, evoking a sociable, adaptable bird often found around human settlements.

\paragraph{Multimodal Fusion} 
The three modalities are fused into a single vector using a learnable attention-weighted sum:
\[
c_{\text{final}} = \alpha_1 \cdot c_{\text{spec}} + \alpha_2 \cdot c_{\text{label}} + \alpha_3 \cdot c_{\text{text}}
\]
where ${\alpha_1, \alpha_2, \alpha_3}$ are learnable weights normalized via softmax such that $\sum_i \alpha_i = 1$. This allows the model to dynamically attend to the most informative cues depending on the categories or training context.

While MFCCs themselves encode some class-specific information, explicitly incorporating categories labels improves training stability and enhances the precision of categories-level generation. This form of redundant but complementary conditioning is particularly effective in low-resource or noisy bioacoustic domains.

During inference, the model generates bird calls by conditioning on user-specified inputs, such as the text description \textit{``Quail morning call''}. Starting from Gaussian noise, the generator iteratively denoises the signal under multimodal conditioning, ultimately producing a waveform that reflects both the target species and the desired acoustic context.                         

\subsection{Evaluation Methodology}  \label{EvaluationMethodology}
To assess the quality of the audio generated by BirdDiff, we adopt a set of objective evaluation metrics. Unlike tasks such as speech or music generation, subjective evaluation (e.g., Mean Opinion Score, MOS) is less applicable to bird call synthesis due to the limited ability of human listeners—especially non-experts—to accurately judge the similarity or authenticity of bird vocalizations. Consequently, we focus exclusively on quantitative and reproducible evaluation techniques.

Specifically, we employ the following metrics: SNR and Itakura–Saito Distance (ISD)~\cite{gray1980distortion,basseville1989distance} for evaluating waveform quality; and Fréchet Audio Distance (FAD)~\cite{kilgour2018fr}, Jensen–Shannon Divergence (JSD)~\cite{menendez1997jensen} combined with Number of Statistically-Different Bins (NDB)~\cite{richardson2018gans}, and classification accuracy  using a pre-trained ResNet50 model~\cite{kong2019acoustic} for evaluating the generated audio distribution and identity preservation.

\paragraph{Baseline}  
As a baseline, we use the original audio dataset without any augmentation or preprocessing. A DiffWave model trained on this unenhanced data serves as our primary point of comparison.

 \paragraph{SNR} 
As introduced in Subsection~\ref{zerothlayer}, we use SegSNR to evaluate changes in signal-to-noise ratio (SNR) before and after enhancement. 
To obtain an initial estimate of the original SNR, we needed to approximate the "signal" and "noise" components from the raw audio without relying on manual annotations. We adopted the following procedure:
\begin{itemize}
\item {Signal Approximation:} A rough estimate of the bird call signal $(x - x_{base})$ was obtained by applying a broad band-pass filter between 2–8\,kHz, which covers the primary frequency range of most bird calls in our dataset.
\item {Noise Approximation:} The background noise was then estimated as the residual, obtained by subtracting the approximated signal from the original waveform $(x - x_{base})$.
\end{itemize}
These components were then used as input to the SegSNR formula to quantify the dataset’s initial  SNR. We chose this automated, frame-based approach because it is well suited to the transient and intermittent nature of bird calls.

\paragraph{FAD}  
FAD measures the distance between the distributions of generated and real audio in an embedding space, providing an approximation of perceived auditory similarity. 
Lower FAD values indicate that the generated audio more closely resembles the real data distribution. This metric is widely used to evaluate generative models in audio tasks.

\paragraph{JSD with NDB}
We extract a 10-dimensional feature vector from each audio sample and compute JSD to assess the similarity between the feature distributions of real and generated samples.  The features include five temporal descriptors—mean amplitude, standard deviation of amplitude, maximum absolute amplitude, mean absolute amplitude, and zero-crossing rate—and five spectral descriptors—mean magnitude, standard deviation of magnitude, normalized dominant frequency position, total spectral energy, and spectral centroid. These were selected because they jointly capture both time-domain dynamics and frequency-domain structure, which are essential for characterizing bird vocalizations. JSD measures the similarity between the distributions of these features for real and generated samples, while NDB complements JSD by quantifying how many histogram bins differ significantly between the two distributions.  
Lower JSD indicates greater similarity between the distributions of generated and real data, while lower NDB values suggest greater sample diversity and reduced mode collapse—both desirable outcomes in generative modeling.

\paragraph{ISD}  
ISD is a frame-wise spectral distance metric that captures differences between power spectra of the original and generated signals. It is particularly sensitive to perceptual distortions in audio, especially in low-SNR scenarios like bird calls. Lower ISD values indicate better reconstruction fidelity at the spectral level and are associated with more realistic waveform synthesis.

\paragraph{Classification Model Evaluation}  
Because classifier training must span all 12 bird-call categories, collecting an entirely separate dataset was not feasible. Therefore, we train a ResNet50-based audio classification model on the original dataset, achieving over 90\% accuracy on both the validation and test sets. This model is then used to classify the generated audio samples.  We evaluated the generated calls from each baseline model, observing progressively higher classification accuracy with our approach.   Higher classification accuracy on generated audio indicates better preservation of category identity and semantic content. 
 It is worth noting that the signal and noise characteristics of the original dataset, as learned by the classifier, may introduce bias into the evaluation due to differences in test data and the classifier’s generalization ability. However, since all evaluations use the same trained classifier, this potential bias is consistently applied and therefore equally reflected across all results.

{\subsection{Experiment Environment}
All experiments were conducted using Python 3.11.13 and CUDA 12.5 on an NVIDIA A100 GPU with 40 GB of RAM. We used PyTorch 2.6.0 and torchaudio 2.6.0 for model development and audio processing, along with NumPy 2.0.2 and SciPy 1.15.3.

\section{Results}\label{results}
We report the results of our experiments in Tables~\ref{tab1}, \ref{tab2} and \ref{tab3}, which encompass both ablation comparisons and per-categories evaluations of the final model.

Table~\ref{tab1} compares our  Multi-band Adaptive Bird-Call Enhancement (MABE) approach in the zeroth layer of BirdDiff  with three classical non-learning-based techniques: Spectral Subtraction~\cite{boll1979spectral}, MMSE-STSA~\cite{ephraim1984stsa}, and MMSE-LSA~\cite{ephraim1985lsa}. These methods have been widely used for speech and audio enhancement due to their simplicity and efficiency, and they remain relevant benchmarks in recent comparative reviews~\cite{lemercier2024review}. 

Our method significantly outperforms the baselines, achieving an average SegSNR improvement of +10.45 dB—substantially higher than the traditional approaches. More importantly, it yields a remarkably low ISD value of 0.54, indicating superior preservation of spectral structure and reduced distortion. All results are averaged across the dataset.

\begin{table}[htbp]
\centering
\caption{Comparison of Non-Learning Audio Enhancement Methods}
\begin{tabular}{|c|c|c|}
\hline
\textbf{Method} & \textbf{SNR Improvement} & \textbf{ISD (↓)} \\
\hline
Spectral Subtraction & +2.36 dB & 1.14 \\
MMSE-STSA &+1.96 dB & 7.25 \\
MMSE-LSA & +2.66 dB & 6.26 \\
\textbf{Ours (MABE)} & \textbf{+10.45 dB} & \textbf{0.54} \\
\hline
\multicolumn{3}{l}{$^{\mathrm{a}}$Lower ISD values indicate lower spectral distortion.}
\end{tabular}
\label{tab1}
\end{table}

\begin{table}[htbp]
\centering
\caption{Generation Quality Under Different Settings}
\begin{tabular}{|c|c|c|c|c|}
\hline
\textbf{Model} & \textbf{\textit{FAD}} & \textbf{\textit{JSD}} & \textbf{\textit{NDB}} & \textbf{\textit{Accuracy}} \\
\hline
DiffWave (Unenhanced) & 0.590 & 0.259 & 7.33 & 35.87\% \\
DiffWave + MABE & 0.281 & 0.227 & 7.00 & 55.57\% \\
BirdDiff & \textbf{0.213} & \textbf{0.226} & \textbf{5.58} & \textbf{70.10\%} \\
\hline
\multicolumn{5}{l}{$^{\mathrm{a}}$All values are averaged over 12 bird categories.}
\end{tabular}
\label{tab2}
\end{table}

Table~\ref{tab2} presents the ablation results comparing   three models. Adaptive enhancement alone reduces FAD by 52.4\% (from 0.590 to 0.281), and yields modest reductions in  JSD and NDB (12.4\% and 4.5\%, respectively), indicating improved alignment with the real data distribution and increased diversity. More importantly, classification accuracy rises from 35.87\% to 55.57\%, highlighting clearer categories-level acoustic features in the generated audio.

Adding multimodal conditioning—via categories labels and text descriptions—further improves performance across most metrics. Our final model achieves the lowest FAD (0.213), a further 24.2\% improvement over adaptive enhancement alone. While JSD changes marginally, NDB decreases substantially from 7.00 to 5.58, suggesting reduced mode collapse. Classification accuracy improves to 70.10\%, demonstrating that the multimodal model successfully generates semantically rich, categories-distinguishable calls.

Table~\ref{tab3} shows a comparison between bird calls generated by our BirdDiff model and those produced using traditional augmentation methods (including mix-up, noise overlay, pitch shifting, and time stretching).  The comparison follows two key principles: (1) all augmentation steps were applied randomly and fully automated, without human intervention, to ensure fairness rather than optimization of enhancement; and (2) the evaluation metrics and classifier are identical to those used in Table~\ref{tab2}, with the classifier trained solely on the original dataset.

The results in Table~\ref{tab3} reveal an interesting yet intuitive pattern. Traditional augmentation methods operate within the original audio distribution; they do not generate new distributions but instead alter existing samples. Consequently, they achieve a very low JSD score (0.070), indicating near-identical global distribution compared to the original dataset. Moreover, the higher NDB score (12) suggests that augmentations—such as added noise—can affect the distribution in fine-grained intervals, underscoring the need for careful manual selection of augmentation strategies.
In contrast, the audio generated by our model shows a higher JSD score (0.287), reflecting a global distribution that differs from the original and introduces greater diversity. At the same time, the lower NDB score (5) indicates that the detailed structure of bird calls is closer to the real distribution. Furthermore, FAD (0.209 vs. 0.217) and classification accuracy (68.65\% vs. 48.34\%) further demonstrate that our model surpasses traditional augmentation methods in both fidelity and semantic identifiability, offering a promising new approach for expanding bioacoustic datasets.

We also examine the impact of BirdDiff across individual bird categories.  Table~\ref{tab3} provides a breakdown of performance by categories using our final model. Accuracy scores vary considerably across categories, from 32.00\% for Common Buzzard to 88.89\% for Quail. High-performing categories such as Quail and  Redshank also exhibit low FAD, JSD, and NDB scores, suggesting strong agreement between perceived quality and semantic correctness.

Interestingly, a decoupling effect is observed in certain categories. For example, Common Buzzard and Western Water Rail yield relatively competitive FAD scores (0.234 and 0.177, respectively), but low classification accuracies (32.00\% and 45.28\%). This highlights the limitation of using FAD alone to evaluate categories-level correctness. Conversely, Grey Heron achieves high accuracy (76.56\%) despite having the highest FAD (0.586), likely due to unique but recognizable vocal structures.

\begin{table}[htbp]
\centering
\caption{Per-categories Evaluation of Generated Bird Calls}
\begin{tabular}{|c|c|c|c|c|c|c|c|c|}
\hline
 & \multicolumn{4}{c|}{\textbf{BirdDiff}} & \multicolumn{4}{c|}{\textbf{Traditional Augmentation}} \\
\hline
\textbf{categories}& \textbf{\textit{FAD}} & \textbf{\textit{JSD}} & \textbf{\textit{NDB}} & \textbf{\textit{Accuracy}} & \textbf{\textit{FAD}} & \textbf{\textit{JSD}} & \textbf{\textit{NDB}} & \textbf{\textit{Accuracy}} \\
\hline
Mallard & 0.264 & 0.256 & 6 & 79.59\% & 0.196 & 0.055 & 8 & 68.67\% \\
Teal & 0.189 & 0.303 & 5 & 75.00\% & 0.161 & 0.060 & 10 & 55.98\% \\
Quail & \textbf{0.134} & 0.273 & 4 & \textbf{88.89\%} & 0.237 & 0.079 & 13 & 40.79\% \\
Pheasant & 0.107 & 0.357 & 7 & 78.85\% & 0.193 & 0.061 & 13 & 53.32\% \\
Red-throated Diver & 0.180 & 0.208 & 7 & 79.66\% & 0.190 & 0.076 & 10 & 40.36\% \\
Grey Heron & 0.586 & 0.433 & 6 & 76.56\% & 0.362 & 0.077 & 9 & 27.53\% \\
Common Buzzard & 0.234 & 0.253 & 6 & 32.00\% & 0.229 & 0.093 & 15 & 54.43\% \\
Western Water Rail & 0.177 & 0.241 & 8 & 45.28\% & 0.230 & 0.084 & 15 & 49.85\% \\
Woodcock & 0.137 & 0.195 & 5 & 53.85\% & 0.177 & 0.043 & 12 & 52.61\% \\
Bar-tailed Godwit & 0.320 & 0.412 & 6 & 70.83\% & 0.209 & 0.075 & 13 & 52.96\% \\
Redshank & 0.185 & \textbf{0.158} & \textbf{3} & 85.00\% & 0.211 & 0.071 & 15 & 49.24\% \\
Sparrow & 0.136 & 0.355 & 4 & 58.33\% & 0.214 & 0.068 & 13 & 34.44\% \\
\hline
Average Results& \textbf{0.209} & 0.287 & \textbf{5} & \textbf{68.65\%} & 0.217 & \textbf{0.070} & 12 & 48.34\% \\
\hline
\multicolumn{9}{l}{$^{\mathrm{*}}$Each result is based on 100 randomly selected samples per category, averaged across 5 runs.}
\end{tabular}
\label{tab3}
\end{table}

In addition to quantitative results, we also provide spectrogram visualizations to offer an intuitive view of the generated audio. Because our dataset contains 12 bird categories, we present only two representative examples—Redshank and Woodcock—due to space constraints. Figures~\ref{fig:redshank} and \ref{fig:woodcock} show these comparisons.

\begin{figure}[htbp]
\centerline{\includegraphics[width=\columnwidth]{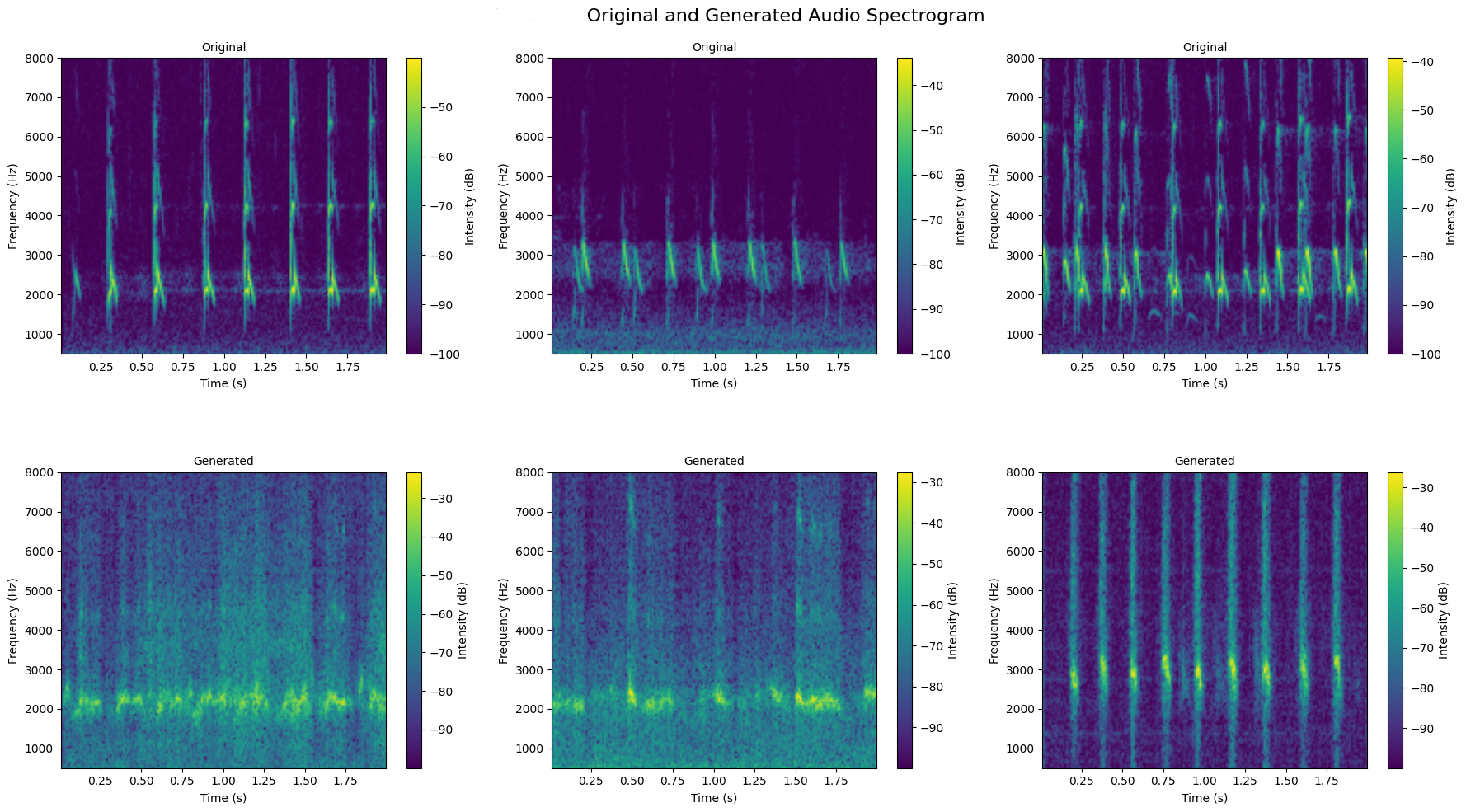}}
\caption{Spectral comparison of original and generated Redshank calls. The first row shows spectrograms of three original recordings, while the second row presents the corresponding generated calls.}
\label{fig:redshank}
\end{figure}
Figure~\ref{fig:redshank} compares the original and generated Redshank calls. Both spectrograms clearly show that the call energy is concentrated between 2000–3000 Hz. The characteristic “continuous jumping” temporal pattern of Redshank vocalizations is also well preserved in the generated samples, illustrating our model’s ability to reproduce species-specific features.

Figure~\ref{fig:woodcock} presents the Woodcock example. Unlike Redshank, Woodcock calls span a broader frequency range (1000–7000 Hz), with multiple harmonics and longer pauses between calls. The generated audio successfully reproduces these distinctive traits. Taken together, the Redshank and Woodcock examples highlight that our model is capable of capturing and replicating the unique acoustic signatures of different species.

\begin{figure}[H]
\centerline{\includegraphics[width=\columnwidth]{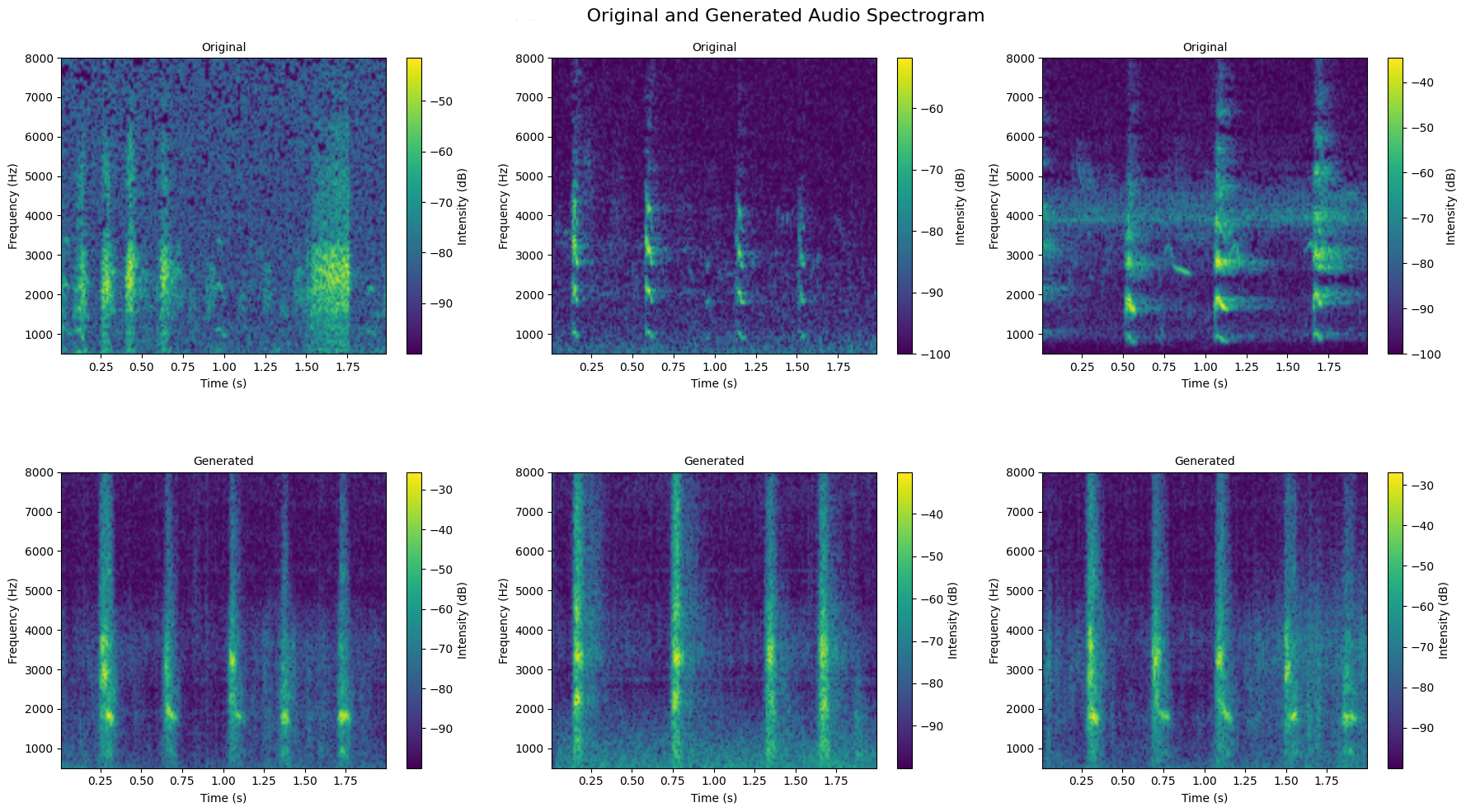}}
\caption{Spectral comparison of original and generated Woodcock calls. The first row shows spectrograms of three original recordings, while the second row presents the corresponding generated calls.}
\label{fig:woodcock}
\end{figure}

\section{Conclusions} \label{Conclusion}
In this study, we propose an adaptive enhancement framework tailored for bird call audio, coupled with a multi-modal conditional diffusion model for waveform generation. Our approach demonstrates substantial improvements across multiple evaluation metrics through step-by-step ablation and comparative experiments. Unlike existing methods that often face trade-offs between audio fidelity and diversity, our method effectively balances both. A core contribution of this work is the ability to generate categories-specific bird calls by incorporating class labels and textual descriptions as conditioning inputs. This multimodal design enables controllable audio synthesis across bird categories while preserving spectral fidelity.

Critically, we find that the success of our generative model depends heavily on the adaptive enhancement strategy. Without preprocessing to improve the signal-to-noise ratio and preserve spectral characteristics, the diffusion model fails to generate intelligible or categories-specific bird calls. The enhancement step is thus essential not only for improving audio quality but also for enabling meaningful categories-level differentiation in the generated outputs.

Nonetheless, several limitations remain. Our dataset includes only 12 bird categories, representing a small fraction of global avian diversity. Additionally, the audio samples are limited to short-duration clips, whereas longer and more structurally complex bird calls may require temporal modeling mechanisms beyond the current framework. Furthermore, the dataset lacks variation in acoustic environments—such as habitat types or weather conditions—which are known to influence vocal behavior in the wild.  We also observe that, in some cases, the generated calls resemble their conditioning spectrograms, suggesting a limitation in diversity that may be addressed through further architectural refinements. In addition, excessively strong spectral guidance can sometimes produce audio that is overly constrained in both time and frequency domains. Enhancing the effectiveness of text prompts in guiding the differentiation of generated audio also remains an open challenge that we have not yet addressed. Finally, our current evaluation relies on ResNet50 classifier trained on the original dataset, which, while informative, does not fully address species-identity preservation. Stronger validation—such as independent classifier training or expert human annotation—remains an important direction for future work.

Future research can build upon our work by incorporating more diverse and ecologically realistic datasets, exploring cross-category generation strategies for rare or endangered birds, and integrating environmental context into the generation process.  We also aim to investigate improved evaluation pipelines that combine independent datasets, alternative classifier training strategies, and expert annotation to more robustly assess the fidelity of species-specific vocal characteristics. We envision that continued advances in generative audio modeling—combined with ecological domain knowledge—will open new opportunities for species monitoring, biodiversity research, and wildlife conservation.

\backmatter


\end{document}